\documentclass[amstex,fleqn,thmsa,a4,12pt]{article}
\usepackage{amsfonts}

\usepackage{amsmath}

\input{tcilatex}
\input tcilatex

\begin{document}

\begin{titlepage}
\setcounter{page}{1}

\title{POINCARE ALGEBRA AND SPACE-TIME CRITICAL DIMENSIONS FOR PARASPINNING 
STRINGS%
\thanks{%
This work was supported by the Algerian Ministry of Education and Research
under contract No.D2501/14/2000.}}
\author{N.BELALOUI\thanks{E-mail : n.belaloui@wissal.dz}, AND H.BENNACER \\
%EndAName
LPMPS, D\'{e}partement de Physique, Facult\'{e} des Sciences,\\
Universit\'{e} Mentouri Constantine, \\
Constantine, Algeria.}
\date{October 26, 2002}
\maketitle

\begin{abstract}
In this paper, we paraquantize the spinning string theory in the Neuveu-Shwarz model . Both
the center of mass variables and the excitation modes of the string verify
paracommutation relations. Except the $\left[p^{\mu },p^{\nu }\right]$
commutator, the two other commutators of Poincar\'e algebra are satisfied.
With the sole use of trilinear relations we find existence
possibilities of spinning strings at space-time dimensions other than $D=10$.
\end{abstract}

\end{titlepage}

\setcounter{page}{2}

\section{Introduction}

In the parabosonic string case [1], a critical study of the Poincar\'{e}
algebra was done and space-time critical dimensions D as functions of the
order of the paraquantization Q were obtained. This work consist in doing an
extention of all these questions to the paraspinning string case in the
Neuveu-Shwarz model. To set the notations, we begin with a brief summary of
some familiar results in spinning string theory [2] [3] [4].

The action is postulated as :

\begin{equation}
S=-\frac{1}{2\pi }\int d\sigma d\tau \left( \partial _{a}X^{\mu }\partial
^{a}X_{\mu }-\bar{\imath}\overline{\psi }^{\mu }\rho ^{a}\partial _{a}\psi
_{\mu }\right)   \tag{1}
\end{equation}
with 
\begin{equation*}
\psi ^{\mu }=\left( 
\begin{array}{c}
\psi _{0}^{\mu } \\ 
\psi _{1}^{\mu }
\end{array}
\right) \quad \quad ;\quad \quad \overline{\psi }=\psi ^{\mu T}\rho ^{0}
\end{equation*}
and 
\begin{equation*}
\rho ^{0}=\left( 
\begin{array}{cc}
0 & -\imath  \\ 
\imath  & 0
\end{array}
\right) \quad \quad ;\quad \quad \rho ^{1}=\left( 
\begin{array}{cc}
0 & \imath  \\ 
\imath  & 0
\end{array}
\right) \quad \quad ;\quad \quad \left\{ \rho ^{a},\rho ^{b}\right\} =-2\eta
^{ab}
\end{equation*}
The solutions are : 
\begin{equation}
X^{\mu }\left( \sigma ,\tau \right) =x^{\mu }+p^{\mu }\tau +\sum_{n\neq 0}%
\displaystyle\frac{1}{n}\alpha _{n}^{\mu }\left( 0\right) \exp \left(
-\imath n\tau \right) \cos n\sigma   \tag{2}
\end{equation}
where $x^{\mu }$ and $p^{\mu }$ are respectively the ''center of mass''
coordinates and the total energy momentum of the string. In the
Neuveu-Shwarz model : 
\begin{equation}
\left\{ 
\begin{array}{c}
\psi _{0}^{\mu }\left( \sigma ,\tau \right) =\displaystyle\frac{1}{\sqrt{2}}%
\sum_{r\in \left( Z+\frac{1}{2}\right) }b_{r}^{\mu }\exp \left[ -\imath
r\left( \tau -\sigma \right) \right]  \\ 
\psi _{1}^{\mu }\left( \sigma ,\tau \right) =\displaystyle\frac{1}{\sqrt{2}}%
\sum_{r\in \left( Z+\frac{1}{2}\right) }b_{r}^{\mu }\exp \left[ -\imath
r\left( \tau +\sigma \right) \right] 
\end{array}
\right.   \tag{3}
\end{equation}
The total angular momentum is given by : 
\begin{equation}
M^{\mu \nu }=M_{0}^{\mu \nu }\left( x\right) +K^{\mu \nu }  \tag{4}
\end{equation}
where $M_{0}^{\mu \nu }\left( x\right) $ is the bosonic part given by : 
\begin{equation}
M_{0}^{\mu \nu }\left( x\right) =x^{\mu }p^{\nu }-x^{\nu }p^{\mu }-\imath
\sum_{n=1}^{\infty }\frac{1}{n}\left( \alpha _{-n}^{\mu }\alpha _{n}^{\nu
}-\alpha _{-n}^{\nu }\alpha _{n}^{\mu }\right)   \tag{5}
\end{equation}
and

\begin{equation}
K^{\mu \nu }=-\frac{\imath }{2}\sum_{r=-\infty }^{\infty }\left( b_{-r}^{\mu
},b_{r}^{\nu }-b_{-r}^{\nu },b_{r}^{\mu }\right)  \tag{6}
\end{equation}

In the same way as in the parabosonic case, a first study of the paraquantum
Poincar\'{e} algebra was done by F.Ardalan and F.Mansouri [5]. These authors
paraquantize the excitation modes of the string and impose to the center of
mass variables to satisfy the ordinary quantum \ commutation relations. This
is done by the choice of a specific direction in the paraspace of the Green
components [6], [7], [8], which requires relative paracommutation relations
between the center of mass \ variables and the excitation modes of the
string. In order that the theory is Poincar\'{e} invariant, the space-time
dimension $D$ and the order of the paraquantization $Q$ are related by the
relation $D=2+\frac{8}{Q}$. Here, we investigate the paraquantization of the
spinning string theory in the Neuveu-shwarz model without the Ardalan and
Mansouri hypothesis on the center of mass variables [5]. We construct the
paraspinning string formalism and we discuss the three commutators of the
Poincar\'{e} algebra. We find that, except the $\left[ p^{\mu },p^{\nu }%
\right] $ commutator, the other results are the same as in the Ardalan and
Mansouri work [5]. In particular, the relation between the space-time
dimensions $D$ and the order of the paraquantization $Q$ is again $D=2+\frac{%
8}{Q}$.

\section{Paraquantum formalism of spinning string}

\subsection{Covariant gauge}

The paraquantization of the theory is carried out by reinterpreting the
classical dynamical variables $\alpha _{n}^{\mu },\,p^{\mu },\,x^{\mu }$ and 
$b_{r}^{\mu }$ as operators satisfying the so called trilinear commutation
relations :

\begin{eqnarray}
\left[ x^{\mu },\left[ p^{\nu },A\right] _{+}\right] &=&2\imath g^{\mu \nu }A
\TCItag{7-a} \\
\left[ x^{\mu },\left[ p^{\nu },p^{\rho }\right] _{+}\right] &=&2\imath
\left( g^{\mu \nu }p^{\rho }+g^{\mu \rho }p^{\nu }\right)  \TCItag{7-b} \\
\left[ \alpha _{n}^{\mu },\left[ \alpha _{m}^{\nu },\alpha _{l}^{\rho }%
\right] _{+}\right] &=&2\left( g^{\mu \nu }n\delta _{n+m,0}\alpha _{l}^{\rho
}+g^{\mu \rho }n\delta _{n+l,0}\alpha _{m}^{\nu }\right)  \TCItag{7-c} \\
\left[ \alpha _{n}^{\mu },\left[ \alpha _{m}^{\nu },B\right] _{+}\right]
&=&2ng^{\mu \nu }\delta _{n+m,0}B  \TCItag{7-d} \\
\left[ b_{r}^{\mu },\left[ b_{s}^{\nu },b_{q}^{\rho }\right] _{-}\right]
&=&2\left( g^{\mu \nu }\delta _{r+s,0}b_{q}^{\rho }-g^{\mu \rho }\delta
_{r+q,0}b_{s}^{\nu }\right)  \TCItag{7-e} \\
\left[ b_{r}^{\mu },\left[ b_{s}^{\nu },C\right] _{+}\right] &=&2g^{\mu \nu
}\delta _{r+s,0}C  \TCItag{7-f}
\end{eqnarray}
and all the other commutators are null. Here $l,n\in Z$ and $r,s,q\in \left( 
\mathbb{Z}+\displaystyle\frac{1}{2}\right) $ and A, B, and C represent the
following operators :

A=$\alpha _{n}^{\rho }\,,\,x^{\rho }$ or $b_{r}^{\rho }$

B= $p^{\rho }\,,\,x^{\rho }$ or $b_{r}^{\rho }$

C= $p^{\rho }\,,\,x^{\rho }$ or $\alpha _{n}^{\rho }$

\subsection{\ Transverse gauge}

In this gauge, the paraquantum operators $x^{-},\,p^{+},\,x^{i},\,p^{i}$ ,$%
\alpha _{n}^{i}$ and $b_{r}^{i}$\ verify the trilinear relations : 
\begin{eqnarray}
\left[ b_{r}^{i},\left[ b_{s}^{j},b_{q}^{k}\right] _{-}\right] &=&2\left(
\delta ^{ij}\delta _{r+s}b_{q}^{k}-\delta ^{ik}\delta _{r+q}b_{s}^{j}\right)
\TCItag{8-a} \\
\left[ \alpha _{n}^{i},\left[ \alpha _{m}^{j},\alpha _{l}^{k}\right] _{+}%
\right] &=&2(\delta ^{ij}n\delta _{n+m,0}\alpha _{l}^{k}+n\delta ^{ik}\delta
_{n+l,0}\alpha _{m}^{j})  \TCItag{8-b} \\
\left[ x^{i},\left[ p^{j},pk\right] _{+}\right] &=&2i\left( \delta
^{ij}pk+\delta ^{ik}p^{j}\right)  \TCItag{8-c} \\
\left[ \alpha _{n}^{i},\left[ \alpha _{m}^{j},D\right] _{+}\right]
&=&2\delta ^{ij}n\delta _{n+m}D  \TCItag{8-d} \\
\left[ b_{r}^{i},\left[ b_{s}^{j},E\right] _{+}\right] &=&2\delta
^{ij}\delta _{r+s}E  \TCItag{8-e} \\
\left[ x^{i},\left[ p^{j},F\right] _{+}\right] &=&2\imath \delta ^{ij}F 
\TCItag{8-f} \\
\left[ x^{-},\left[ p^{+},G\right] _{+}\right] &=&2\imath G  \TCItag{8-g}
\end{eqnarray}
{\Large \ }and all the others commutators are null. Here D, E, F, and G
represent the following operators $:$

D = $x^{-},\,p^{+},\,x^{k},\,p^{k}$ or $b_{q}^{k}.$

E =\ $x^{-},\,p^{+},\,x^{k},\,p^{k}$ or $\alpha _{n}^{k}.$

F = $x^{-},\,p^{+},\,x^{k},\,$ $\alpha _{n}^{k}$ or $b_{r}^{k}$

G = $x^{-},\,\,x^{k},\,p^{k}$ $\alpha _{n}^{k}$ or $b_{r}^{k}$

\section{Paraquantum Poincar\'{e} Algebra}

In view of the form of the relations (7), the quantum form of the Poincar%
\'{e} algebra generators $M^{\mu \nu }$(4,5,6) presents an order ambiguity
problem so that there must bee rewriten on the basis of an adequate
symmetrisation which takes the form (4) where now : 
\begin{equation}
M_{0}^{\mu \nu }\left( x\right) =l^{\mu \nu }+E^{\mu \nu }  \tag{9}
\end{equation}
with 
\begin{equation}
l^{\mu \nu }=\frac{1}{2}\left[ x^{\mu },p^{\nu }\right] _{+}-\frac{1}{2}%
\left[ x^{\nu },p^{\mu }\right] _{+}  \tag{10}
\end{equation}
and 
\begin{equation}
E^{\mu \nu }=-\frac{\imath }{2}\sum_{n=1}^{\infty }\frac{1}{n}\left( \left[
\alpha _{-n}^{\mu },\alpha _{n}^{\nu }\right] _{+}-\left[ \alpha _{-n}^{\nu
},\alpha _{n}^{\mu }\right] _{+}\right)   \tag{11}
\end{equation}
and

\begin{equation}
K^{\mu \nu }=-\frac{\imath }{4}\sum_{r=-\infty }^{\infty }\left( \left[
b_{-r}^{\mu },b_{r}^{\nu }\right] _{-}-\left[ b_{-r}^{\nu },b_{r}^{\mu }%
\right] _{-}\right)  \tag{12}
\end{equation}

With a direct application of the trilinear relations (7), let us perform the
second and the third commutators of the algebra : 
\begin{equation}
\left[ p^{\mu },M^{\nu \rho }\right] =\left[ p^{\mu },M_{0}^{\nu \rho }%
\right] +\left[ p^{\mu },K^{\nu \rho }\right]   \tag{13}
\end{equation}
By the use of (7-a,b), it is easy to see that : 
\begin{equation}
\left[ p^{\mu },K^{\nu \rho }\right] =0  \tag{14}
\end{equation}
and

\begin{equation}
\left[ p^{\mu },M_{0}^{\nu \rho }\right] =\imath \left( g^{\mu \rho }p^{\nu
}-g^{\mu \nu }p^{\rho }\right)  \tag{15}
\end{equation}
so that 
\begin{equation}
\left[ p^{\mu },M^{\nu \rho }\right] =\imath \left( g^{\mu \rho }p^{\nu
}-g^{\mu \nu }p^{\rho }\right)  \tag{16}
\end{equation}

Now, for the third commutator

\begin{equation}
\left[ M^{\mu \nu },M^{\alpha \beta }\right] =\left[ M_{0}^{\mu \nu }\left(
x\right) ,M_{0}^{\alpha \beta }\left( x\right) \right] +\left[ M_{0}^{\mu
\nu }\left( x\right) ,K^{\alpha \beta }\right] +\left[ K^{\mu \nu
},M_{0}^{\alpha \beta }\left( x\right) \right] +\left[ K^{\mu \nu
},K^{\alpha \beta }\right]  \tag{17}
\end{equation}
the first term is given by [1] : 
\begin{equation}
\left[ M_{0}^{\mu \nu },M_{0}^{\rho \sigma }\right] =\imath g^{\nu \rho
}M_{0}^{\sigma \mu }-\imath g^{\mu \sigma }M_{0}^{\nu \rho }-\imath g^{\nu
\sigma }M_{0}^{\rho \mu }+\imath g^{\mu \rho }M_{0}^{\nu \sigma }  \tag{18}
\end{equation}

It is again clear from (7) that : 
\begin{equation}
\left[ M_{0}^{\mu \nu }\left( x\right) ,K^{\alpha \beta }\right] =\left[
K^{\mu \nu },M_{0}^{\alpha \beta }\left( x\right) \right] =0  \tag{19}
\end{equation}

\bigskip Let us now consider the commutator : 
\begin{multline}
\left[ K^{\mu \nu },K^{\alpha \beta }\right] =-\frac{1}{16}\sum_{r,s}\left\{ %
\left[ \left[ b_{-r}^{\mu },b_{r}^{\nu }\right] ,\left[ b_{-s}^{\alpha
},b_{s}^{\beta }\right] \right] -\left( \mu ,\nu ,\alpha \leftrightarrow
\beta \right) \right.  \notag \\
\left. -\left( \mu \leftrightarrow \nu ,\alpha ,\beta \right) +\left( \mu
\leftrightarrow \nu ,\alpha \leftrightarrow \beta \right) \right\}  \tag{20}
\end{multline}
The first term gives : 
\begin{eqnarray}
A &=&\sum_{r,s}\left[ \left[ b_{-r}^{\mu },b_{r}^{\nu }\right] ,\left[
b_{-s}^{\alpha },b_{s}^{\beta }\right] \right]  \TCItag{21} \\
&=&\sum_{r,s}\left\{ b_{-r}^{\mu }\left[ b_{r}^{\nu },\left[ b_{-s}^{\alpha
},b_{s}^{\beta }\right] \right] +\left[ b_{-r}^{\mu },\left[ b_{-s}^{\alpha
},b_{s}^{\beta }\right] \right] b_{r}^{\nu }-b_{r}^{\nu }\left[ b_{-r}^{\mu
},\left[ b_{-s}^{\alpha },b_{s}^{\beta }\right] \right] -\left[ b_{r}^{\nu },%
\left[ b_{-s}^{\alpha },b_{s}^{\beta }\right] \right] b_{-r}^{\mu }\right\} 
\notag
\end{eqnarray}
With the use of (7-e), (21) becomes : 
\begin{equation}
A=2\sum_{r=-\infty }^{+\infty }\left( g^{\nu \alpha }\left[ b_{-r}^{\mu
},b_{r}^{\beta }\right] -g^{\mu \beta }\left[ b_{-r}^{\alpha },b_{r}^{\nu }%
\right] -g^{\nu \beta }\left[ b_{-r}^{\mu },b_{r}^{\alpha }\right] +g^{\mu
\alpha }\left[ b_{-r}^{\beta },b_{r}^{\nu }\right] \right)  \tag{22}
\end{equation}
In the same way, one can perform the other terms of (20) and obtain :

\begin{multline}
\left[ K^{\mu \nu },K^{\alpha \beta }\right] =\imath \sum_{r=-\infty
}^{+\infty }\left\{ -\frac{\imath }{4}\left\{ g^{\nu \alpha }\left( \left[
b_{-r}^{\beta },b_{r}^{\mu }\right] -\left[ b_{-r}^{\mu },b_{r}^{\beta }%
\right] \right) -g^{\mu \beta }\left( \left[ b_{-r}^{\alpha },b_{r}^{\nu }%
\right] -\left[ b_{-r}^{\nu },b_{r}^{\alpha }\right] \right) \right. \right.
\notag \\
\quad \quad \quad \quad \quad \left. \left. -g^{\nu \beta }\left( \left[
b_{-r}^{\alpha },b_{r}^{\mu }\right] -\left[ b_{-r}^{\mu },b_{r}^{\alpha }%
\right] \right) +g^{\mu \alpha }\left( \left[ b_{-r}^{\nu },b_{r}^{\beta }%
\right] -\left[ b_{-r}^{\beta },b_{r}^{\nu }\right] \right) \right\} \right\}
\tag{23}
\end{multline}
then

\begin{equation}
\left[ K^{\mu \nu },K^{\alpha \beta }\right] =\imath \left( g^{\nu \alpha
}K^{\beta \mu }-g^{\mu \beta }K^{\alpha \nu }-g^{\nu \beta }K^{\alpha \mu
}+g^{\mu \alpha }K^{\nu \beta }\right)  \tag{24}
\end{equation}
When combined with (18) and (19), (17) gives : 
\begin{equation}
\left[ M^{\mu \nu },M^{\alpha \beta }\right] =\imath \left( g^{\nu \alpha
}M^{\beta \mu }-g^{\mu \beta }M^{\alpha \nu }-g^{\nu \beta }M^{\alpha \mu
}+g^{\mu \alpha }M^{\nu \beta }\right)  \tag{25}
\end{equation}

Now, for the first commutator $\left[ p^{\mu },p^{\nu }\right] $ of the
algebra, one can only write $\left[ p^{\mu },\left[ p^{\nu },p^{\sigma }%
\right] _{+}\right] =0$ and not $\left[ p^{\mu },p^{\nu }\right] =0!$

\section{Space-time critical dimensions}

As in the ordinary case, one can obtain the space-time critical dimension by
performing, in the transverse gauge, the commutator $\left[ M^{i-},M^{j-}%
\right] .$

Let us introduce, in the transverse gauge, the generators $M^{i-}$ in the
form : 
\begin{equation}
M^{i-}=M_{0}^{i-}\left( x\right) +K^{i-}  \tag{26}
\end{equation}
where 
\begin{equation}
M_{0}^{i-}\left( x\right) =l^{i-}+E^{i-}  \tag{27}
\end{equation}
\begin{eqnarray}
l^{i-} &=&\frac{1}{2}\left[ x^{i},\frac{1}{p^{+}}\right] _{+}\alpha _{0}^{-}-%
\frac{1}{2}\left[ x^{-},p^{i}\right] _{+}  \TCItag{28} \\
E^{i-} &=&-\frac{\imath }{2}\sum_{n=1}^{\infty }\frac{1}{n}\left( \left[
\alpha _{-n}^{i},\frac{1}{p^{+}}\right] _{+}\alpha _{n}^{-}-\alpha _{-n}^{-}%
\left[ \alpha _{n}^{i},\frac{1}{p^{+}}\right] _{+}\right)  \TCItag{29}
\end{eqnarray}
with 
\begin{equation}
\alpha _{-n}^{-}=-\frac{1}{4}\sum_{l=-\infty }^{+\infty }\left[ \alpha
_{n-l}^{i},\alpha _{l}^{i}\right] _{+}-\frac{1}{4}\sum_{r=-\infty }^{+\infty
}\left( r-\frac{n}{2}\right) \left[ b_{n-r}^{i},b_{r}^{i}\right] -\frac{a}{2}%
\delta _{n,0}  \tag{30}
\end{equation}
and

\begin{equation}
K^{i-}=-\frac{\imath }{4}\sum_{r=\frac{1}{2}}^{\infty }\left( \left[
b_{-r}^{i},\frac{1}{p^{+}}\right] _{+}G_{r}-G_{-r}\left[ b_{r}^{i},\frac{1}{%
p^{+}}\right] _{+}\right)  \tag{31}
\end{equation}
with 
\begin{equation}
G_{r}=\frac{1}{2}\sum_{n=-\infty }^{+\infty }\left[ \alpha _{n}^{i},b_{r}^{i}%
\right] _{+}  \tag{32}
\end{equation}

Let us now perform this commutator : 
\begin{equation}
\left[ M^{i-},M^{j-}\right] =\left[ M_{0}^{i-},M_{0}^{j-}\right] +\left[
M_{0}^{i-},K^{j-}\right] +\left[ K^{i-},M_{0}^{j-}\right] +\left[
K^{i-},K^{j-}\right]  \tag{33}
\end{equation}
Projecting the equation $\left[ M^{i-},M^{j-}\right] =0$ on the physical
states $\alpha _{-m}^{k}\left| 0\right\rangle $ and $b_{-s}^{k}\left|
0\right\rangle $, one can write : 
\begin{equation}
\left\langle 0\right| \alpha _{m}^{l}\left[ M^{i-},M^{j-}\right] \alpha
_{-m}^{k}\left| 0\right\rangle +\left\langle 0\right| b_{s}^{l}\left[
M^{i-},M^{j-}\right] b_{-s}^{k}\left| 0\right\rangle =0  \tag{34}
\end{equation}
We begin by computing the first commutator of (33). First, notice that one
can write : 
\begin{equation}
\left\langle 0\right| b_{s}^{l}\left[ M_{0}^{i-},M_{0}^{j-}\right]
b_{-s}^{k}\left| 0\right\rangle \equiv 0  \tag{35}
\end{equation}
For the other mean value on the physical states $\alpha _{-m}^{k}\left|
0\right\rangle $, except the term which is done by [9] :

\begin{eqnarray}
C_{4} &=&\frac{1}{4}\sum_{n,n^{\prime }}\left\langle 0\right| \alpha _{m}^{l}%
\left[ \frac{\alpha _{-n}^{i}}{n},\frac{1}{p^{+}}\right] _{+}\alpha
_{-n}^{-}\alpha _{-n^{\prime }}^{-}\left[ \frac{\alpha _{-n^{\prime }}^{j}}{%
n^{\prime }},\frac{1}{p^{+}}\right] _{+}\alpha _{-m}^{k}\left|
0\right\rangle   \notag \\
&=&\frac{1}{\left( p^{+}\right) ^{2}}\delta ^{li}\delta ^{jk}\left( Q\frac{%
D-2}{8}m\left( m^{2}-1\right) +2ma\right)   \TCItag{36}
\end{eqnarray}
all the other terms are analog to parabosonic case [1]. Then, we obtain : 
\begin{multline}
\left[ M_{0}^{i-},M_{0}^{j-}\right] =-\frac{1}{2\left( p^{+}\right) ^{2}}%
\sum_{n=1}^{\infty }\left( \left[ \alpha _{-n}^{i},\alpha _{n}^{j}\right]
_{+}-\left[ \alpha _{-n}^{j},\alpha _{n}^{i}\right] _{+}\right)   \notag \\
\times \left[ -2n+Q\frac{D-2}{8}\left( n-\frac{1}{n}\right) +\frac{2a}{n}%
\right]   \tag{37}
\end{multline}
For the second and the third terms of (33), one can write : 
\begin{equation}
\left[ M_{0}^{i-},K^{j-}\right] +\left[ K^{i-},M_{0}^{j-}\right] =\left[
l^{i-}+E^{i-},K^{j-}\right] -\left( i\leftrightarrow j\right)   \tag{38}
\end{equation}
Before computing the mean value of the commutator $\left[ l^{i-},K^{j-}%
\right] $, we first transform it as follows : we notice that by the use of
the relations (8-a,b,d,e), one can verify that : 
\begin{eqnarray}
\left[ \alpha _{n}^{-},b_{r}^{i}\right]  &=&\left( r+\frac{n}{2}\right)
b_{n+r}^{i}  \TCItag{39} \\
\left[ \alpha _{n}^{-},G_{r}\right]  &=&\left( r-\frac{n}{2}\right) G_{r+n} 
\TCItag{40}
\end{eqnarray}
so that :

\begin{equation}
\left[ \alpha _{0}^{-},\left[ b_{-r}^{j},\frac{1}{p^{+}}\right] _{+}G_{r}%
\right] =-r\left[ b_{-r}^{j},\frac{1}{p^{+}}\right] _{+}G_{r}+\left[ \frac{1%
}{p^{+}},b_{-r}^{j}\right] _{+}rG_{r}=0  \tag{41}
\end{equation}
On the other hand, and again by the use of (8-f,g), one can verify that : 
\begin{eqnarray}
\left[ x^{i},G_{r}\right] &=&\imath b_{r}^{i}  \TCItag{42} \\
\left[ x^{-},\left[ \frac{1}{p^{+}},H\right] _{+}\right] &=&-\frac{2\imath }{%
\left( p^{+}\right) ^{2}}H  \TCItag{43}
\end{eqnarray}
where the operator H = $x^{-},\,\,x^{k},\,p^{k}$ $\alpha _{n}^{k}$ or $%
b_{r}^{k}$. Then we obtain : 
\begin{multline}
\left[ l^{i-},K^{j-}\right] =\frac{1}{4}\sum_{r=\frac{1}{2}}^{\infty }\left( %
\left[ \frac{1}{p^{+}},b_{-r}^{j}\right] _{+}\left[ b_{r}^{i},\frac{1}{p^{+}}%
\right] _{+}-\left[ \frac{1}{p^{+}},b_{-r}^{i}\right] _{+}\left[ \frac{1}{%
p^{+}},b_{r}^{j}\right] _{+}\right) \alpha _{0}^{-}  \notag \\
-\frac{1}{4}\left( \frac{1}{p^{+}}\right) ^{2}\sum_{r=\frac{1}{2}}^{\infty
}\left( \left[ b_{-r}^{j},p^{i}\right] _{+}G_{r}-G_{-r}\left[ b_{r}^{j},p^{i}%
\right] _{+}\right)  \tag{44}
\end{multline}
It is easy to see that : 
\begin{equation}
\left\langle 0\right| \alpha _{m}^{l}\left[ l^{i-},K^{j-}\right] \alpha
_{-m}^{k}\left| 0\right\rangle =0  \tag{45}
\end{equation}
Now, for the other mean value, one can prove that : 
\begin{eqnarray}
\left\langle 0\right| b_{s}^{l}\left[ l^{i-},K^{j-}\right] b_{-s}^{k}\left|
0\right\rangle &=&-\frac{1}{2\left( p^{+}\right) ^{2}}\sum_{r=\frac{1}{2}%
}^{\infty }\left\langle 0\right| b_{s}^{l}r\left( \left[ b_{-r}^{j},b_{r}^{i}%
\right] -\left[ b_{-r}^{i},b_{r}^{j}\right] \right) b_{-s}^{k}\left|
0\right\rangle  \notag \\
&&-\frac{1}{2\left( p^{+}\right) ^{2}}\delta ^{lj}p^{i}p^{k}+\frac{1}{%
2\left( p^{+}\right) ^{2}}\delta ^{jk}p^{i}p^{l}  \TCItag{46}
\end{eqnarray}
Notice that one can again write : 
\begin{equation}
\left\langle 0\right| b_{s}^{l}\left( \left[ l^{i-},K^{j-}\right] +\left[
K^{i-},l^{j-}\right] \right) b_{-s}^{k}\left| 0\right\rangle =2\left\langle
0\right| b_{s}^{l}\left[ l^{i-},K^{j-}\right] b_{-s}^{k}\left| 0\right\rangle
\tag{47}
\end{equation}
In the same way, for the commutator $\left[ E^{i-},K^{j-}\right] $, we
compute the mean value 
\begin{equation}
\left\langle 0\right| \alpha _{m}^{l}\left[ E^{i-},K^{j-}\right] \alpha
_{-m}^{k}\left| 0\right\rangle +\left\langle 0\right| b_{s}^{l}\left[
E^{i-},K^{j-}\right] b_{-s}^{k}\left| 0\right\rangle  \tag{48}
\end{equation}
To do this, we need to perform the following non null expressions, which
give the results

\begin{eqnarray}
H_{1} &=&-\frac{1}{4}\sum_{r=\frac{1}{2}}^{\infty }\sum_{n=1}^{\infty
}\left\langle 0\right| \alpha _{m}^{l}\left[ \frac{1}{p^{+}},\alpha _{-n}^{i}%
\right] _{+}\alpha _{n}^{-}\left[ \frac{1}{p^{+}},b_{-r}^{j}\right]
_{+}G_{r}\alpha _{-m}^{k}.\left| 0\right\rangle  \notag \\
&=&\frac{3m^{3}}{4\left( p^{+}\right) ^{2}}\delta ^{li}\delta ^{jk} 
\TCItag{49}
\end{eqnarray}

\begin{eqnarray}
H_{2} &=&\frac{1}{4}\sum_{r,n}\left\langle 0\right| b_{s}^{l}\left[ \left[ 
\frac{\alpha _{-n}^{i}}{n},\frac{1}{p^{+}}\right] _{+}\alpha _{n}^{-},G_{-r}%
\left[ b_{r}^{j},\frac{1}{p^{+}}\right] _{+}\right] b_{-s}^{k}\left|
0\right\rangle  \notag \\
&=&\delta ^{jk}\delta ^{li}\frac{1}{2\left( p^{+}\right) ^{2}}\left( s^{2}-%
\frac{1}{4}+s\right) +\frac{1}{\left( p^{+}\right) ^{2}}\delta
^{jk}p^{l}p^{i}  \TCItag{50}
\end{eqnarray}
\begin{eqnarray}
H_{3} &=&-\frac{1}{4}\sum_{r,n}\left\langle 0\right| b_{s}^{l}\alpha
_{-n}^{-}\left[ \frac{\alpha _{n}^{i}}{n},\frac{1}{p^{+}}\right] _{+}\left[
b_{-n}^{j},\frac{1}{p^{+}}\right] _{+}G_{r}b_{-s}^{k}\left| 0\right\rangle 
\notag \\
&=&\frac{1}{2\left( p^{+}\right) ^{2}}\delta ^{lj}\delta ^{ik}\left( s^{2}-%
\frac{1}{4}\right)  \TCItag{51}
\end{eqnarray}
It follows that : 
\begin{multline}
\left\langle 0\right| \alpha _{m}^{l}\left[ E^{i-},K^{j-}\right] \alpha
_{-m}^{k}\left| 0\right\rangle +\left\langle 0\right| b_{s}^{l}\left[
E^{i-},K^{j-}\right] b_{-s}^{k}\left| 0\right\rangle =  \notag \\
-\frac{1}{2\left( p^{+}\right) ^{2}}\left\{ \left( \frac{3}{4}m^{3}-s^{2}+%
\frac{1}{4}\right) \left( \delta ^{li}\delta ^{jk}-\delta ^{lj}\delta
^{ik}\right) +\frac{1}{2}\delta ^{jk}\delta ^{li}\left( s^{2}-\frac{1}{4}%
+s\right) +\delta ^{jk}p^{l}p^{i}\right\}  \tag{52}
\end{multline}
Then we can write the result : 
\begin{multline}
\left\langle 0\right| \alpha _{m}^{l}\left( \left[ E^{i-},K^{j-}\right] +%
\left[ K^{i-},E^{j-}\right] \right) \alpha _{-m}^{k}\left| 0\right\rangle
+\left\langle 0\right| b_{s}^{l}\left( \left[ E^{i-},K^{j-}\right] +\left[
K^{i-},E^{j-}\right] \right) b_{-s}^{k}\left| 0\right\rangle =  \notag \\
\frac{1}{\left( p^{+}\right) ^{2}}\left\{ \left( \frac{3}{4}m^{3}-\frac{s^{2}%
}{2}+\frac{1}{8}+\frac{s}{2}\right) \left( \delta ^{li}\delta ^{jk}-\delta
^{lj}\delta ^{ik}\right) +\delta ^{jk}p^{l}p^{i}-\delta
^{ik}p^{l}p^{j}\right\}  \tag{53}
\end{multline}

Lastly, one can perform the mean value of the last commutator of (33) and
obtain (Appendix A) : 
\begin{eqnarray}
&&\left\langle 0\right| b_{s}^{l}\left[ K^{i-},K^{j-}\right]
b_{-s}^{k}\left| 0\right\rangle +\left\langle 0\right| \alpha _{m}^{l}\left[
K^{i-},K^{j-}\right] \alpha _{-m}^{k}\left| 0\right\rangle   \notag \\
&=&\frac{1}{4\left( p^{+}\right) ^{2}}\left( \delta ^{li}\delta ^{jk}-\delta
^{lj}\delta ^{ik}\right) \left[ \left( s^{2}-\frac{1}{4}\right) +s-\frac{Q}{2%
}\left( D-2\right) \left( s^{2}-\frac{1}{4}\right) -8a-m^{3}\right]   \notag
\\
&&\quad \quad \quad \quad \quad \quad \quad +\frac{1}{4\left( p^{+}\right)
^{2}}\left( \delta ^{li}p^{j}p^{k}-\delta ^{lj}p^{i}p^{k}+\delta
^{jk}p^{l}p^{i}-\delta ^{ik}p^{l}p^{j}\right)   \TCItag{54}
\end{eqnarray}

Now by regrouping all these results, in terms of operators, one obtain :

\begin{multline}
\left[ M^{i-},M^{j-}\right] =\frac{1}{2\left( p^{+}\right) ^{2}}%
\sum_{n=1}^{\infty }\left( \left[ \alpha _{-n}^{i},\alpha _{n}^{j}\right]
_{+}-\left[ \alpha _{-n}^{j},\alpha _{n}^{i}\right] _{+}\right) \left( Q%
\frac{D-2}{8}\left( n-\frac{1}{n}\right) +\frac{2a}{n}-n\right)  \notag \\
-\frac{1}{2\left( p^{+}\right) ^{2}}\sum_{r=\frac{1}{2}}^{\infty }\left( %
\left[ b_{-r}^{i},b_{r}^{j}\right] _{-}-\left[ b_{-r}^{j},b_{r}^{i}\right]
_{-}\right) \left( \left( Q\frac{D-2}{8}-1\right) \left( r^{2}-\frac{1}{4}%
\right) +2a-1\right)  \tag{55}
\end{multline}

We are thus led to conclude that, in order to have $\left[ M^{i-},M^{j-}%
\right] =0$ , one must have :

\begin{equation}
\left\{ 
\begin{array}{l}
D=2+\frac{8}{Q} \\ 
a=\frac{1}{2}
\end{array}
\right.  \tag{56}
\end{equation}

\section{Conclusion}

This work consist in paraquantizing the spinning string by imposing
paracommutation relations to the classical variables $X^{\mu }\left( \sigma
,\tau \right) $, $P^{\mu }\left( \sigma ,\tau \right) $ and $\psi _{A}^{\mu
}\left( \sigma ,\tau \right) $. Unlike in Ardalan and Mansouri work [5],
this requires that both the center of mass variables and the excitation
modes of the string verify paracommutation relations.To satisfy this, one
must have $\left[ X^{\mu (\alpha )}\left( \sigma ,\tau \right) ,P^{\nu
(\alpha )}(\sigma ^{\prime },\tau )\right] =$ $\imath g^{\mu \nu }\delta
(\sigma -\sigma ^{\prime })$ [1], But the Ardalan \ and Mansouri hypothesis,
characterized by the anzatz $x^{\mu (\beta )}=x^{\mu }\delta _{\beta 1}$and $%
p^{\mu (\beta )}=p^{\mu }\delta _{\beta 1}$, leads to the result $\left[
X^{\mu (\alpha )}\left( \sigma ,\tau \right) ,P^{\nu (\alpha )}(\sigma
^{\prime },\tau )\right] =$ $\imath g^{\mu \nu }\left[ \delta (\sigma
-\sigma ^{\prime })-(1-\delta _{\alpha 1})\right] $, which is different from
the latter. With the only use of the trilinear relations (7), one can prove
that : 
\begin{equation*}
\left[ p^{\mu },M^{\nu \rho }\right] =-\imath g^{\mu \nu }p^{\rho }+\imath
g^{\mu \rho }p^{\nu }
\end{equation*}
\begin{equation*}
\left[ M^{\mu \nu },M^{\rho \sigma }\right] =\imath g^{\nu \rho }M^{\sigma
\mu }-\imath g^{\mu \sigma }M^{\nu \rho }-\imath g^{\nu \sigma }M^{\rho \mu
}+\imath g^{\mu \rho }M^{\nu \sigma }
\end{equation*}

In the transverse gauge, in order to have $\left[ M^{i-},M^{j-}\right] =0$,
the space-time critical dimension $D$ is calculated with the only use of the
trilinear relations (8). Like in Ardalan and Mansouri work [5], $D$ is again
given as function of the paraquantization order through the relation $D=2+%
\frac{8}{Q}$. Thus, one can have paraspinning strings with critical
dimensions $D=10,6,4,3$ (respectively in orders $Q=1,2,4,8$). This coincide
with the dimensions in which fractional superstrings can be formulated [10],
[11]. \ \ \ \ \ \ \ \ \ \ \ \ \ \ \ \ \ \ \ \ \ \ \ \ \ \ \ \ \ \ \ \ \ \ \
\ \ \ \ \ \ \ \ \ \ \ 

Some questions arises; in order to satisfy $($ $\left[ p^{\mu },p^{\nu }%
\right] =0)$, one adopt the Ardalan and Mansouri anzatz, then the question
is can one speak about the paraquantum formalism on the classical variables $%
X^{\mu }\left( \sigma ,\tau \right) $, $P^{\mu }\left( \sigma ,\tau \right) $
and $\psi _{A}^{\mu }\left( \sigma ,\tau \right) $? On the other hand, if we
paraquantize these classical variables by imposing them to satisfy
paracommutation relations, what we may only write is $\left[ p^{\mu },\left[
p^{\nu },p^{\sigma }\right] _{+}\right] =0$. Then, what about the space-time
properties?

\bigskip \newpage

\textbf{Acknowledgments}

We are pleased to thank professor Keith R.\ Dienes from the Department of
Physics of the University of Arizona for pointing out some works on
fractional superstrings, and for his advises.

{\Large \textbf{Appendix}}

Let us set 
\begin{equation}
\left\langle 0\right| \alpha _{n}^{l}K^{i-}K^{j-}\alpha _{-m}^{k}\left|
0\right\rangle +\left\langle 0\right| b_{s}^{l}K^{i-}K^{j-}b_{-s}^{k}\left|
0\right\rangle =\sum_{i=1}^{4}D_{i}  \tag{A-1}
\end{equation}
computing 
\begin{eqnarray}
D_{1} &=&\frac{1}{4}\sum_{r,r^{\prime }}\left\langle 0\right| b_{s}^{l}\left[
b_{-r}^{i},\frac{1}{p^{+}}\right] _{+}G_{r}\left[ b_{-r}^{j},\frac{1}{p^{+}}%
\right] _{+}G_{r^{\prime }}b_{-s}^{k}\left| 0\right\rangle  \notag \\
&=&\frac{1}{\left( p^{+}\right) ^{2}}\delta ^{li}\delta ^{jk}\frac{1}{2}%
\left[ \left( s^{2}-\frac{1}{4}\right) +s\right] +\frac{1}{\left(
p^{+}\right) ^{2}}\delta ^{li}p^{j}p^{k}  \TCItag{A-2}
\end{eqnarray}
\begin{eqnarray}
D_{2} &=&-\frac{1}{4}\sum_{r,r^{\prime }}\left\langle 0\right| b_{s}^{l}%
\left[ b_{-r}^{i},\frac{1}{p^{+}}\right] _{+}G_{r}G_{-r^{\prime }}\left[
b_{r^{\prime }}^{j},\frac{1}{p^{+}}\right] _{+}b_{-s}^{k}\left|
0\right\rangle  \notag \\
&=&-\frac{1}{\left( p^{+}\right) ^{2}}\delta ^{li}\delta ^{jk}\left[ \frac{Q%
}{2}\left( D-2\right) \left( s^{2}-\frac{1}{4}\right) +2a\right] 
\TCItag{A-3}
\end{eqnarray}
\begin{eqnarray}
D_{3} &=&-\frac{1}{4}\sum_{r,r^{\prime }}\left\langle 0\right| \alpha
_{m}^{l}G_{-r}\left[ b_{r}^{i},\frac{1}{p^{+}}\right] _{+}\left[
b_{-r^{\prime }}^{j},\frac{1}{p^{+}}\right] _{+}G_{r^{\prime }}\alpha
_{-m}^{k}\left| 0\right\rangle  \notag \\
&=&\frac{m^{3}}{\left( p^{+}\right) ^{2}}\delta ^{lj}\delta ^{ik} 
\TCItag{A-4}
\end{eqnarray}
\begin{eqnarray}
D_{4} &=&\frac{1}{4}\sum_{r,r^{\prime }}\left\langle 0\right| b_{s}^{l}G_{-r}%
\left[ \frac{1}{p^{+}},b_{r}^{i}\right] _{+}G_{-r^{\prime }}\left[
b_{r^{\prime }}^{j},\frac{1}{p^{+}}\right] _{+}b_{-s}^{k}\left|
0\right\rangle  \notag \\
&=&\frac{1}{2\left( p^{+}\right) ^{2}}\delta ^{jk}\delta ^{li}\frac{1}{2}%
\left[ \left( s^{2}-\frac{1}{4}\right) +s\right] +\frac{1}{\left(
p^{+}\right) ^{2}}\delta ^{jk}p^{l}p^{i}  \TCItag{A-5}
\end{eqnarray}
\bigskip it is then straightforward to obtain the relation (54).

\bigskip

\bigskip

\bigskip

\bigskip

\bigskip

\

\ 

\end{document}